\documentclass[aps,twocolumn,prb,preprintnumbers,superscriptaddress,showpacs]{revtex4}
\usepackage{graphicx}
\usepackage{dcolumn}
\usepackage{revsymb}
\usepackage{bm}
\usepackage[latin1]{inputenc}

\Roman{section}
\begin{document}
\def\coup{K}
\def\heli{\Upsilon}
\def\be{\begin{eqnarray}}
\def\ee{\end{eqnarray}}
\preprint{none}

\title{Three-dimensional Ising model conf\/ined in low-porosity aerogels:\\ 
a Monte Carlo study} 

\author{Ricardo Paredes V.} 
\affiliation{Centro de F\'{\i}sica, 
Instituto Venezolano de Investigaciones Cient{\'\i}ficas, 
Apartado 21827, Caracas 1020A, Venezuela.} 
\affiliation{Technische Universiteit Delft, DelftChemTech, Particle Technology, 
Julianalaan 136, 2628 BL, Delft, The Netherlands} 
\author{Carlos V\'asquez} 
\affiliation{Departamento de F{\'\i}sica, Universidad Sim\'on Bol{\'\i}var, 
Apartado 89000, Caracas 1080A, Venezuela.}

\begin{abstract}
The influence of correlated impurities on the critical behaviour of the 3D Ising model 
is studied using Monte Carlo simulations. Spins are confined into the pores 
of simulated aerogels (dif\/fusion limited cluster-cluster aggregation) 
in order to study the effect of quenched disorder on the critical behaviour 
of this magnetic system. Finite size scaling is used to estimate critical 
couplings and exponents. Long-range correlated disorder does not affect critical 
behavior. Asymptotic exponents differ from those of the pure 3D Ising model (3DIS), 
but it is impossible, with our precision, to distinguish them from the randomly 
diluted Ising model (RDIS). 
\end{abstract}
\maketitle

\section{Introduction}\label{intro}

The influence of quenched disorder on phase transitions has been studied since long 
time ago now. In 1974, Harris \cite{Harris} established this famous criterion: 
{\em uncorrelated disorder is not relevant, for a second order phase transition, if the specific heat exponent is 
negative} ($\alpha <  0$). 
The criterion was generalized by Weinrib and Halperin (WH) \cite{wh83} for any disorder 
distribution whose correlation function exhibits a power-law tail, {\em i.~e.}, 
$g(r)\sim r^{-a}$ as $r\to \infty$. Disorder is shown to be relevant in these cases:
\be 
d\nu - 2 &<& 0 \mbox{~   if   ~} a \geq d,\label{whsrc}\\ 
a\nu - 2 &<& 0 \mbox{~   if   ~} a < d,\label{whlrc} 
\ee 
being $d$ the dimension, and $\nu$ the correlation length exponent of the pure system. 
After Josephson hyperscaling ($2-d\nu=\alpha$) Harris criterion is recovered in the 
short-range correlated (SRC) regime (\ref{whsrc}). In contrast, the long-range 
correlated (LRC) regime extends the criterion to systems satisfying condition 
(\ref{whlrc}), even if $\alpha < 0$. This generalization explains why critical 
exponents for the superfluid (SF) transition of $^{4}\mbox{He}$ change when the fluid 
is confined in very light aerogels \cite{chan,YoonChanAerogel} and do not when confined 
in, for example, porous gold \cite{YoonChanGold}. Aerogels are fractal for several 
length scales \cite{Vacher}, while porous gold has exponentially decaying correlations 
beyond the size of a typical pore \cite{YoonChanGold}. Nevertheless, authors 
\cite{chan,YoonChanAerogel} argued that the critical behaviour of SF $^{4}\mbox{He}$ 
in aerogels yet poses intriguing questions to be solved.

For instance, light aerogels are fractal for several length scales, up to a certain 
value $\Lambda$ that depends on aerogel density. Beyond this length, the disordered 
structure becomes homogeneous, entering an uncorrelated regime. After Harris criterion, 
as the SF correlation length $\xi$ gets larger than $\Lambda$, disorder should become 
irrelevant, because $\alpha\simeq -0.011$ is negative for this system. 
Yoon {\em et al.}~\cite{YoonChanAerogel} estimated that this crossover should appear at 
$t=|T-T_c|/T_c\simeq 10^{-4}$ but, although they approached $T_c$ as close as 
$t\simeq 10^{-5}$, no crossover to bulk exponents was observed. A different 
universality class was evident for the SF transition of $^4$He, when confined in 
aerogels. An explanation to these changes was given using Monte Carlo (MC) simulations 
of the three-dimensional XY (3DXY) model, confined in aerogel-like structures
\cite{vasquez}. The SF transition belongs to the 3DXY universality class, and 
correlated disorder could be relevant provided that the WH condition (\ref{whlrc}) at 
$r\to \infty$ is fulfilled. V\'asquez {\em et al.}~\cite{vasquez,vasquez3} 
showed that changes occur because of {\em hidden} LRC, inherent to the process of aerogel 
formation. Using simulated aerogels, made by diffusion limited cluster-cluster 
aggregation (DLCA) \cite{meakin,jullien}, authors showed that different LRC subsets are 
physically well defined within the whole aerogel structure. Specifically, 
{\em gelling clusters} (to be defined later in this paper) are shown to be the relevant 
structures defining the critical behaviour of the 3DXY model in DLCA aerogels 
\cite{vasquez}.

In this paper, we study the three-dimensional Ising model (3DIS), in presence of such 
aerogel-like structures. The pure 3DIS has a positive specific heat exponent 
($\alpha \simeq 0.11$), so any type of disorder, correlated or not, will be relevant. 
If Ising spins are collocated in the pores of aerogels, criticality will be affected by 
LRC as well as by SRC disorder. Our main purpose is to elucidate which among these 
effects dominates the critical behaviour of the 3DIS model in this case. Along this 
paper, we report the results of extensive MC simulations of the 3DIS in the pores of 
DLCA aerogels at fixed porosity $\varphi=80\%$, in order to clarify this point. 

The rest of this paper is organized as follows: Section \ref{ante3DIS} is a brief 
review about diluted Ising systems studied in the past. Section \ref{model-sim} is due 
to explain the model first, then the simulation procedure in detail, with a preamble on 
self-averaging, in order to validate our procedure; two methods to obtain accurate 
values of the critical coupling are presented at the end. Thermal and magnetic 
effective exponents are presented, and their asymptotic behaviours are discussed in 
Section \ref{effective}. Finally, section \ref{conclu} is due to present some 
concluding remarks. 

\section{Antecedents on the diluted 3DIS}\label{ante3DIS}
Many experimental, theoretical, and computational works have been done to date, in 
order to study the critical behaviour of the 3DIS model in presence of quenched 
disorder. Most of numerical and theoretical works address the random-diluted 3DIS 
(RDIS), {\em i.~e.}, the Ising model in presence {\em non-correlated} distributions of 
impurities\cite{grinstein,holovatch,folk,vicari,calabreseMC}. 

Appart, concerning disordered LRC structures, Ballesteros and Parisi \cite{ballespar} 
simulated the 3DIS, with dislocations represented by lines of impurities generated at 
random. Correlations for this type of disorder decay with an exponent $a=2$. They 
obtain a correlation length exponent $\nu_{\mbox{\tiny LRC}} \approx 1$ for the impure 
system, thus confirming the result of WH, that this exponent should be \cite{wh83} 
$\nu_{\mbox{\tiny LRC}}=2/a$. Marqu\'es {\em et al.}~\cite{marques} also simulated a 
diluted 3DIS, but with spins located {\em on} LRC sites. These sites were provided by a 
previous simulation of the pure 3DIS model; taking all sites from the cluster of 
dominating spin orientation at $T_c$, these are then occupied by the interacting Ising 
spins to simulate. Clusters for this {\em thermally diluted} Ising system have anomalous 
dimension $\eta_{\mbox{\tiny pure}}\approx 0.03$, which gives 
$a = 2 - \eta_{\mbox{\tiny pure}} \approx 1.97$. They obtain an exponent 
$\nu_{\mbox{\tiny LRC}} \approx 1$, also in agreement with the WH expression. In both 
cases, LRC disorder is relevant for criticality. Nevertheless, this particular result 
from WH has proven recently not to be correct at more accurate approximations. Using 
two-loop expansions, Prudnikov {\em et al.} \cite{prudnikov}, showed that the exponent 
$\nu_{\mbox{\tiny LRC}}$ depends on both, the internal dimension of the order parameter 
$m$ and the exponent $a$, not the case in WH's conjecture, independent of $m$. For both 
systems, Prudnikovs' prediction yields $\nu_{\mbox{\tiny LRC}} \approx 0.72\neq 1$.

Experiments about the critical point of the liquid-vapour (LV) transition of 
$^4\mbox{He}$ and $\mbox{N}_{2}$, confined in $95\%$ porous aerogels 
\cite{wong4He,wongN2}, concern directly the problem we are addressing in this paper. 
Bulk $^4\mbox{He}$ near its LV critical point belongs to the 3DIS universality class, 
and aerogel-like disorder has proven to contain both, LRC and SRC disordered structures 
\cite{vasquez,hasmy2}. Wong {\em et al.} \cite{wong4He,wongN2} report, for the 
order parameter, exponents $\beta = 0.28(5)$ and $0.35(5)$, respectively, consistent 
with that calculated for the pure 3DIS by Guida and Zinn-Justin \cite{Guida} 
$\beta = 0.326(3)$. However, in the same experiments, the specific heat curves present 
finite peaks at $T_c$, characteristic of a negative exponent $\alpha$, definitely 
different from the corresponding $\alpha_{\mbox{\tiny pure}}\approx 0.11$ for the pure 
3DIS. Actually, within error bars, results for $\beta$ are also consistent with the 
corresponding RDIS value \cite{ballesteros}, $\beta = 0.355(5)$. After these 
experimental results one may take non-correlated, SRC instead of LRC disorder, within 
aerogels, to be the relevant one for the critical behaviour of the 3DIS in aerogel 
pores\cite{vasquez3}. 

Renormalization group  (RG) calculations for the 3DIS with weak amounts of disorder 
show that a new universality class appears, different from that of the pure 3DIS 
\cite{holovatch,folk,newman,jug,mayer,leguillou,calabreseRG}, and consistent with 
Harris criterion. Since D.~P.~Landau \cite{landau}, using MC simulations, concluded 
that the exponents for the 3DIS with random impurities differ from those of the pure 
system, different works stated that exponents depend on the concentration of 
impurities. Until 1990, when Heuer \cite{heuer1,heuer2} began to clarify that 
differences with RG calculations are due to the fact that exponents obtained from 
simulations were basically effective ones, and not the asymptotic ones. 

Ballesteros {\em et al.}~\cite{ballesteros}, using a $p$-reweighting method in MC 
simulations, found the exponents for the RDIS universality class to be independent 
from the concentration of impurities $p$. This was confirmed for the random bond Ising 
model in $d=3$ by Berche {\em et al.}~\cite{peberche}, looking at finite size scaling for 
the critical temperature. All those calculations motivated further MC and RG studies 
about the crossover between the effective and really asymptotic critical behaviour 
\cite{calabreseRG,peberche,calabreseMC}.

Definitive evidences of a new universality class were obtained using neutron scattering 
in the antiferromagnets $\mbox{Mn}_{1-x}\mbox{Zn}_x\mbox{F}_2$ \cite{belanger} and 
$\mbox{Fe}_{x}\mbox{Zn}_{1-x}\mbox{F}_2$ \cite{mitchell}. For the first system, 
exponents $\nu = 0.70(2)$ and $\gamma = 1.37(4)$ are obtained, while $\nu = 0.69(2)$ 
and $\gamma = 1.31(3)$, are the results for the second one. These results clearly 
differ from bulk exponents, see for instance those calculated by Guida and Zinn-Justin 
\cite{Guida}: $\nu=0.6304(13),\quad \gamma=1.239(5)$. 

The critical behaviour of magnetic systems confined in aerogel--like structures, may be 
subject of competing LRC and SRC influences. The 3DXY model in pores of DLCA aerogels, 
for instance, presents new exponents due only to the presence of gelling clusters, 
which are LRC, while the SRC components are irrelevant to the transition\cite{vasquez}. 
However, for the Ising model under the same kind of confinement, two different effects 
may be present. Simulations under strictly LRC types of disorder 
\cite{ballespar,marques}, give exponents consistent with the result of Weinrib and 
Halperin. On the other hand, experimental results about the critical point of LV 
transitions in aerogels \cite{wong4He,wongN2} point to the relevance of the 
uncorrelated part of disorder. 

\section{The model and simulation procedure}\label{model-sim} 
The 3DIS in the presence of impurities, on a simple cubic lattice with nearest-neighbor 
interaction, is described by the Hamiltonian: 
\be 
\beta\mathcal{H}=-\frac{J}{k_{\mbox{\tiny B}}T} 
\sum_{\langle i j\rangle} \epsilon_i \epsilon_j s_i s_j, 
\label{hamilton} 
\ee 
where $s=\pm 1$ are the spin variables, $k_{\mbox{\tiny B}}$ is the Boltzmann constant, 
and $J$ is the coupling. In what follows, $k_{\mbox{\tiny B}}=1$ and $T=1$. The sets 
$\{\epsilon_i\}$ represent quenched variables chosen to be $0$ if the site is an 
impurity and $1$ if the site is occupied by a spin. These sets of impurities are taken 
randomly (RDIS) or from DLCA aerogels (AEIS). 

\subsection{Disorder generation and MC simulations}\label{dlcaproc} 
At the beginning, sites are occupied by a uniform random distribution of $N$ particles, 
so their volume fraction is $c=N/L^3$. To simulate the RDIS, this initial distribution 
of disorder is held through the rest of the MC simulation. 

Instead, for AEIS simulations to take place, aerogels are generated through the 
on-lattice DLCA algorithm \cite{meakin,jullien} with periodic boundary conditions 
(PBC). Monomers and clusters diffuse randomly with diffusivity constants $D$ which 
depend on their mass $n$ through $D\sim n^{-1/d_f}$. The fractal dimension $d_f$ has 
been taken equal to its value in three dimensions\cite{hasmy1}, $d_f\approx 1.8$. They 
stick irreversibly when they come in contact, and then the process follows up until a 
single cluster is obtained. This model is known to reproduce well the geometric 
features of real aerogels \cite{estrDLCA}. 

At a given stage in this DLCA process, the first aggregate to reach opposite sides of 
the simulation box in any direction is called the gelling cluster (GC). It has been 
shown\cite{vasquez,hasmy2} that the correlation function for this GC is algebraic up to 
a cutoff, which diverges as $L\to\infty$. In other words, these objects are fractal 
(LRC). Right after the GC is built, many other smaller clusters ({\em islands}) continue 
to diffuse, and finally attach themselves to the GC at {\em random} sites. The resulting 
DLCA cluster (GC with islands) becomes homogeneous at a very small cutoff, in spite of 
the existence of a physically well defined fractal structure, the GC. This cutoff 
increases as the concentration decreases, a feature already observed for real silica 
aerogels\cite{Vacher}. At the volume fraction employed in the present work, $c=0.2$, 
the cutoff is so small (a few lattice constants) that DLCA clusters must be considered 
as non-fractals. Thus, the presence of islands, which represent the SRC subset within 
the whole DLCA cluster, actually {\em hide} the LRC behaviour of gelling clusters. It is 
in this sense that aerogel-like structures must be considered as a mixture of LRC and 
SRC disorder distributions. 

Disordered samples are generated by the procedures described above. A MC simulation is 
performed for each sample of interacting 3DIS spins, placed at empty sites left by 
impurities. Physical observables, denoted by caligraphs $\mathcal{O}$, are calculated 
at each indepedent MC step, and then corresponding ensemble (thermal) averages 
$\langle\mathcal{O}\rangle$ are taken over the MC time-series. Wolff algorithm 
\cite{Wolff} is used to update spins. In disordered systems, this algorithm tends to 
prevent some regions from being visited by growing Wolff clusters. If the concentration 
of impurities is small, this problem can be solved by adding some Metropolis updates 
along the simulation process \cite{ballesteros}. We chose this method and include some 
Metropolis sweeps to shake all spins, after a fixed number of Wolff steps. An 
independent step is taken after one correlation time $\tau$, which has been estimated 
from preliminary simulations. After enough steps for thermalization, a fixed number 
($N_{T}=1000$) of independent MC steps are performed to calculate thermal averages. 
Equivalent simulations take place for ${N}_{S}=2000$ different samples and, finally, 
averages over disorder are taken $O=[{\langle \mathcal{O}\rangle}]$ (denoted by square 
brackets). System sizes are $L=8,12,16,24,32,48,64,96$ for the RDIS, and 
$L=8,12,16,24,28,32,40,48,56,64,80,96$ for the AEIS. 

\subsection{Measured observables.}
The magnetization (order parameter) is calculated by 
\be
\mathcal{M}=\frac{1}{N}\sum_i^{L^3} \epsilon_i s_i, 
\ee  
where $N=c{L^3}$ is the total number of spins. Thermal averages 
$\langle\mathcal{M}\rangle$ are taken, and averages over disorder 
$M(J)=\left[\langle \mathcal{M}\rangle\right](J)$, are then calculated after the former 
have been extrapolated by reweighting \cite{ferreswend}. The procedure is described 
below in detail. In terms of the magnetization, we define the susceptibility as: 
\be 
\chi = J L^3 
\left[\langle{\mathcal{M}^2}\rangle-\langle\mathcal{M}\rangle^2\right]. 
\label{defchi} 
\ee 

The energy is correspondingly defined by 
\be\label{energy}
\mathcal{E} = -J\sum_{\langle i j\rangle} 
\epsilon_i \epsilon_j s_i s_j,\; E = \left[\langle\mathcal{E}\rangle\right]
\ee
and then the specific heat is obtained from fluctuations of the energy: 
\be\label{suscep}
c_h= L^{-3}\left[\langle{\mathcal{E}^2}
\rangle-\langle\mathcal{E}\rangle^2\right].
\ee

Logarithmic derivatives of $n^{th}$ moments $M^n$ of the magnetization ($n=1,2,4$), 
respect to the coupling, are calculated through the energy-magnetization covariance: 
\be\label{logderiv} 
\left[{\frac{\partial\ln \langle\mathcal{M}^n\rangle}{\partial J}}\right]= 
-\left[\frac{\langle \mathcal{M}^n\mathcal{E}\rangle-\langle\mathcal{E}\rangle\langle\mathcal{M}^n\rangle} 
{\langle \mathcal{M}^n\rangle}\right] 
\ee 

\subsection{Simulation temperatures and reweighting.} 
V\'asquez {\em et al.}~\cite{vasquez} found the phase diagram for the 3DXY model in the 
pores of DLCA aerogels. They obtained $T_c(c)/T_c(0)=J_c(0)/J_c(c)$ as a function of 
the concentration of impurities $c$, being $J_c(c)$ the 3DXY critical coupling at 
volume fraction $c$ of the aerogel. The shape of this phase diagram comes basically 
from the porous structure of disorder, specially at low concentrations. Using this 
information and the critical coupling $J_c(0)=0.2216595(26)$ for the pure 3DIS
\cite{ferrenberg}, V\'asquez \cite{vasquez3} made a rough estimate for the critical 
coupling for the 3DIS in the pores of DLCA aerogels at $c=0.2$. Making simulations at 
this rough estimate, and using lattice sizes $L=10-80$ and finite size scaling, the 
value $J_c(0.2)=0.25855(3)$ is obtained for the critical coupling. Although for those 
simulations the number of disorder realizations is low ($N_{\mbox{\tiny S}}=30$), they 
obtain critical exponents close enough to those reported for the RDIS. 

In the present work, all simulations were done at $c=0.2$, using simulation 
temperatures $\widetilde{J_c}=0.285745$ for the RDIS (following Calabrese 
{\em et al.}~\cite{calabreseMC}), and at the above estimate $\widetilde{J_c}=0.25855$ for 
the AEIS. Physical quantities at $J\simeq\widetilde{J_c}$, are obtained by the 
reweighting method introduced by Ferrenberg and Swendsen \cite{ferreswend}. This 
procedure was used for each disorder realization at each system size $L$. Each 
thermodynamic quantity was then averaged over disorder for each $J$ within the 
extrapolation interval. Finally, maxima of 
$\left[{\chi}(J,L)\right]$, $\left[ {c}(J,L)\right]$ and 
$\left[\partial \ln \langle \mathcal{M}^n \rangle /\partial J(J,L)\right]$ were 
obtained from averaged curves, with their corresponding pseudocritical couplings 
$J_{c}^*(L)$. 

\subsection{Disorder sampling and self-averaging.}
\begin{figure}[ht!]
\includegraphics[width=8cm]{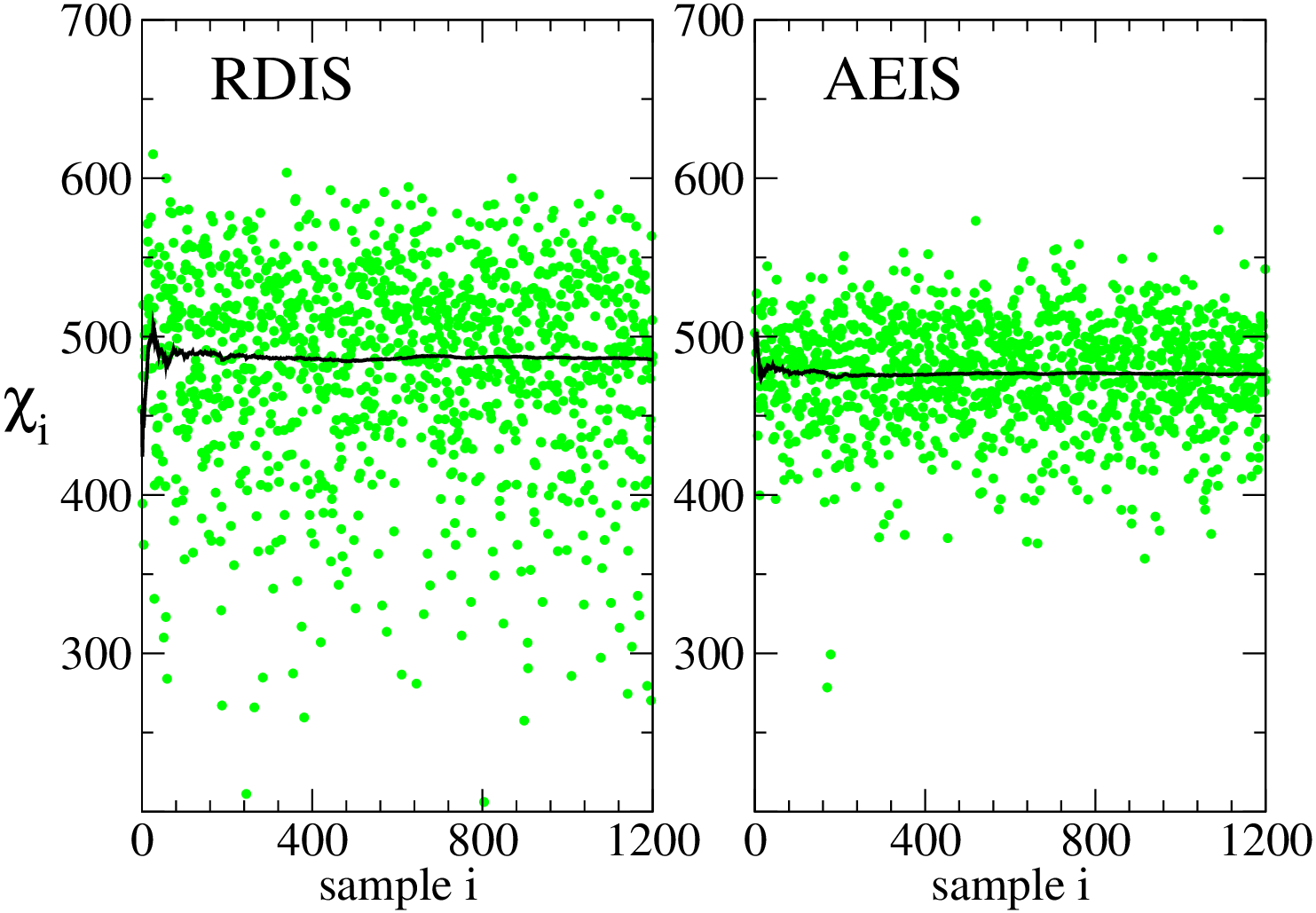}
\caption{(Color online) Distribution of the susceptibility for the RDIS ({\em left}) and 
AEIS ({\em right}) types of disorder, at a concentration of impurities $c=0.2$ and 
lattice size $L=96$. Simulations were performed at $\widetilde{J_c}=0.285745$ for the 
RDIS \cite{calabreseMC}, and at $\widetilde{J_c}=0.25855$ for the AEIS. Running 
averages over the samples $\left[{\chi_i}\right]$ are shown by black thick solid lines.
\label{chi_i-vs-sample_i}}
\end{figure}

\begin{figure}[ht!]
\includegraphics[width=7cm]{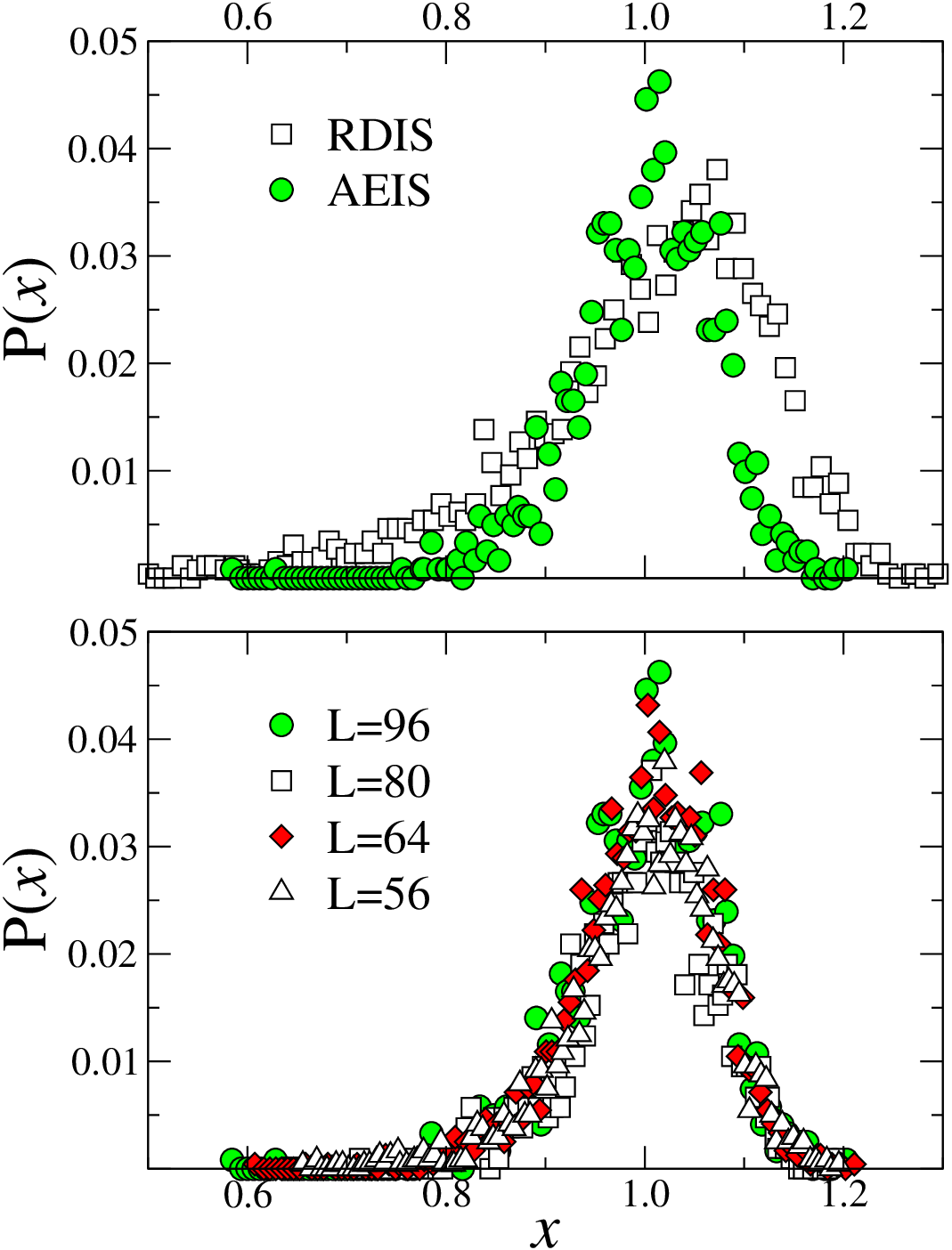}
\caption{(Color online) Probability distribution  of the normalized susceptibility 
$x=\chi_i/\left[{\chi}\right]$. ({\em Top}) Distribution for the AEIS looks sharper and 
more symetrical than that for the RDIS (data taken from Fig.~\ref{chi_i-vs-sample_i}).
({\em Bot.}) Distributions at different lattice sizes for simulations of the AEIS. The 
distribution width appears independent on the lattice size. \label{Pchivschi}} 
\end{figure}

This part is dedicated to determine a suitable number $N_S$ of disorder realizations to 
obtain critical exponents for the AEIS at enough accuracy. A complete study on 
probability distributions for different thermodynamic quantities is performed, and 
results compared with the corresponding ones for the RDIS to decide $N_S$. As an 
example, in Fig.~\ref{chi_i-vs-sample_i} we depict critical susceptibility points, 
obtained for corresponding disorder samples in the RDIS (left) and AEIS (right) cases 
(both at $c=0.2$). Points come from simulations at the couplings $\widetilde{J_c}$ 
estimated above, using the largest lattice sizes ($L=96$). This distribution looks 
sharper in the AEIS case, and notably more symetric respect to the average than in the 
RDIS case. Top of  Fig.~\ref{Pchivschi} shows the probability distribution for the 
susceptibility in both cases. 
\newline
\newline

As remarked, the distribution for AEIS (filled circles) is sharper and 
more symetric than the distribution for RDIS (empty squares). Note that, in terms of 
the normalized variable $x=\chi_i/\left[{\chi}\right]$, the maximum of the distribution 
for AEIS is closer to the average $x=1$ than for RDIS. Bottom of Fig.~\ref{Pchivschi} 
shows the probability distribution of susceptibility for different lattice sizes. 
Independence of the width of distributions from $L$ is clear, typical for systems 
lacking of self-averaging, which is the expected behaviour of thermodynamic quantities 
for any disordered system at criticality\cite{wisemandomany}. 

Self-averaging can be quantitatively checked by the normalized squared width 
\cite{marques2} $R_{A}$: 
\be 
R_{A}(L)=\frac{\left[{A^2(L)}\right]- 
\left[{A(L)}\right]^2}{\left[{A(L)}\right]^2}, 
\ee 
being $A$ any given thermodynamic quantity. In this paper, $R_{A}(L)$ was estimated for 
the RDIS to compare with previously reported values. We obtain, as $L\to \infty$, 
$R_M \to 0.054$ for the magnetization, and $R_\chi\to 0.016$ for the susceptibility, 
both in agreement with previous results \cite{wisemandomany2}. The ratio here obtained 
$R_M/R_\chi\simeq 3.4$ disagrees with RG predictions: Aharony and Harris 
\cite{aharonyharris} obtained, using $\epsilon=4-d$ expansions, that the leading term 
is $R_M/R_\chi=1/4$. The discrepancy may come from higher order terms in the expansion, 
and not from the definition of the susceptibility as was suggested by Berche 
{\em et al.}~\cite{peberche}. Note also that, in the present work, the definition for the 
susceptibility, 
$\chi= JL^3\left[\langle\mathcal{M}^2\rangle-\langle\mathcal{M}\rangle^2\right]$, 
differs from that used by Wiseman and Domany \cite{wisemandomany2}, 
$\chi= JL^3\left[\langle\mathcal{M}^2\rangle\right]$.
\newline
\newline 
\begin{figure}[ht!]
\hspace{-0.5cm}
\includegraphics[width=7cm]{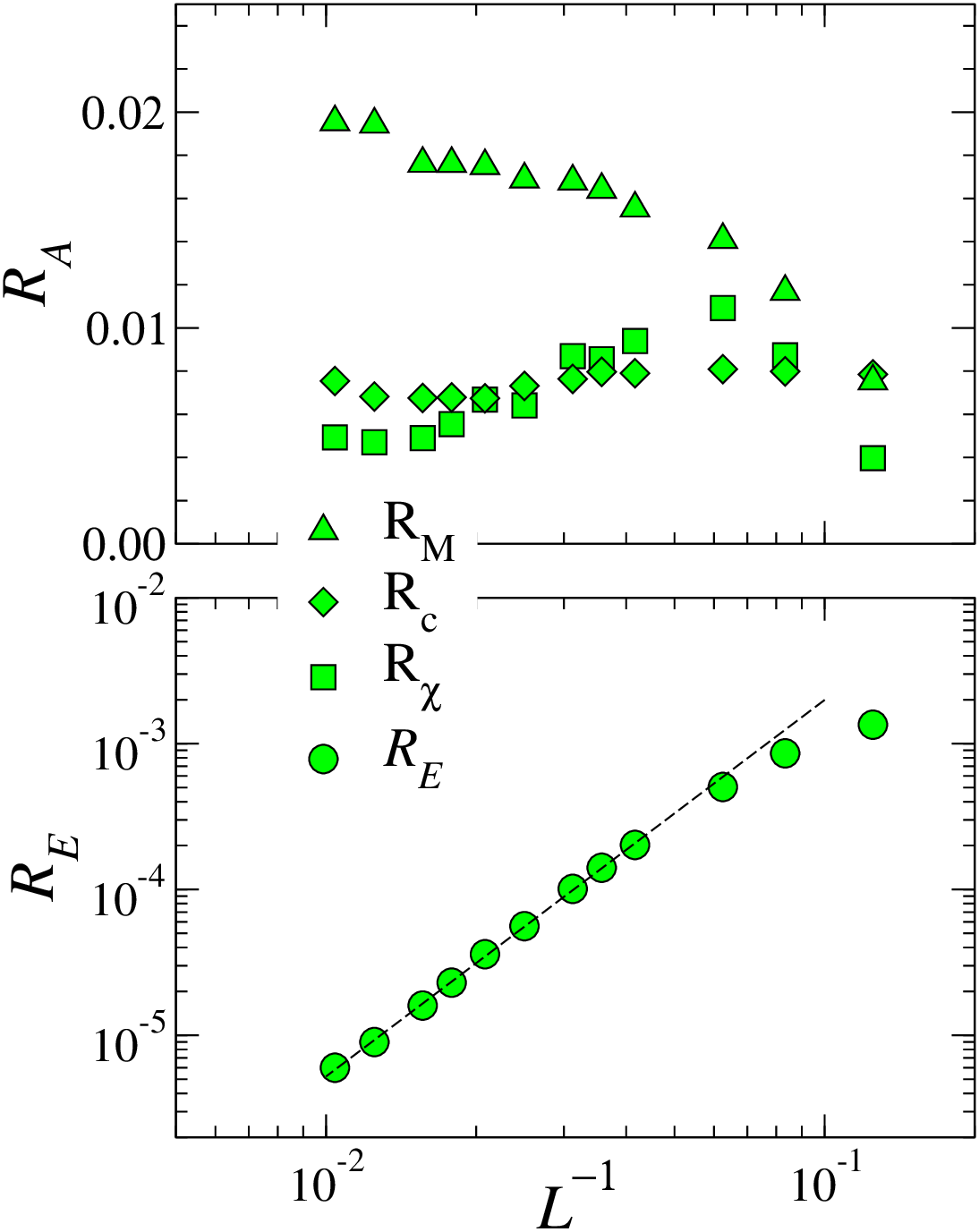}
\caption{(Color online) Normalized squared widths $R_A$ versus inverse system size 
$L^{-1}$ for the AEIS at criticality. ({\em Top}) Asymptotic non-zero values for $R_A$ 
as $L\to\infty$, for the magnetization ($M$), the susceptibility ($\chi$), and the 
specific heat ($c_h$), evidence the lack of self-averaging in these quantities. 
({\em Bottom}) The power-law behaviour $R_E\sim L^{-x}$, with a fitting exponent 
$x\simeq 2.58<d$, indicates that the energy ($E$) is weakly self-averaged.\label{RvsL}}
\end{figure}

Results for $R_{A}(L)$, plotted versus $L^{-1}$, in the AEIS case, are shown on 
Fig.~\ref{RvsL}, being $A$ the order parameter $M$, the susceptibility 
$\chi$, the specific heat $c_h$, and the energy $E$, respectively. As observed for 
$M,\,\chi$ and $c_h$ (top), $R_{A}$ tends to non-zero values as $L\to\infty$, though 
asymptotic limits for $R_A$ are smaller for the AEIS than for the RDIS, as expected 
($R_M\to 0.020$ and $R_\chi\to 0.0028$, as $L\to\infty$). The power-law behaviour 
$R_E \sim L^{-x}$ has been depicted for the energy (bottom), being the fitting exponent 
$x\simeq 2.58$. Thus, the energy is weakly self-averaged\cite{wisemandomany} ($x<d$). 
The same type of behaviour is obtained, in this work, for the energy in the RDIS case, 
in agreement with previously reported results\cite{wisemandomany}. 
According to these analyses, the number $N_{S}$ of disorder realizations, suitable to 
estimate critical exponents, is larger for the RDIS than for the AEIS. In next section, 
we report some values for effective critical exponents for both models. Our results for 
the RDIS agree well with previously published results\cite{wisemandomany2}, thus, the 
same number of realizations for the AEIS will be enough to estimate critical exponents, 
as the values of the normalized squared widths $R_{A}$ are substantially lower than 
those obtained for the RDIS. 

\subsection{Critical couplings}\label{secJc} 
\begin{figure}[ht!]
\includegraphics[width=7cm]{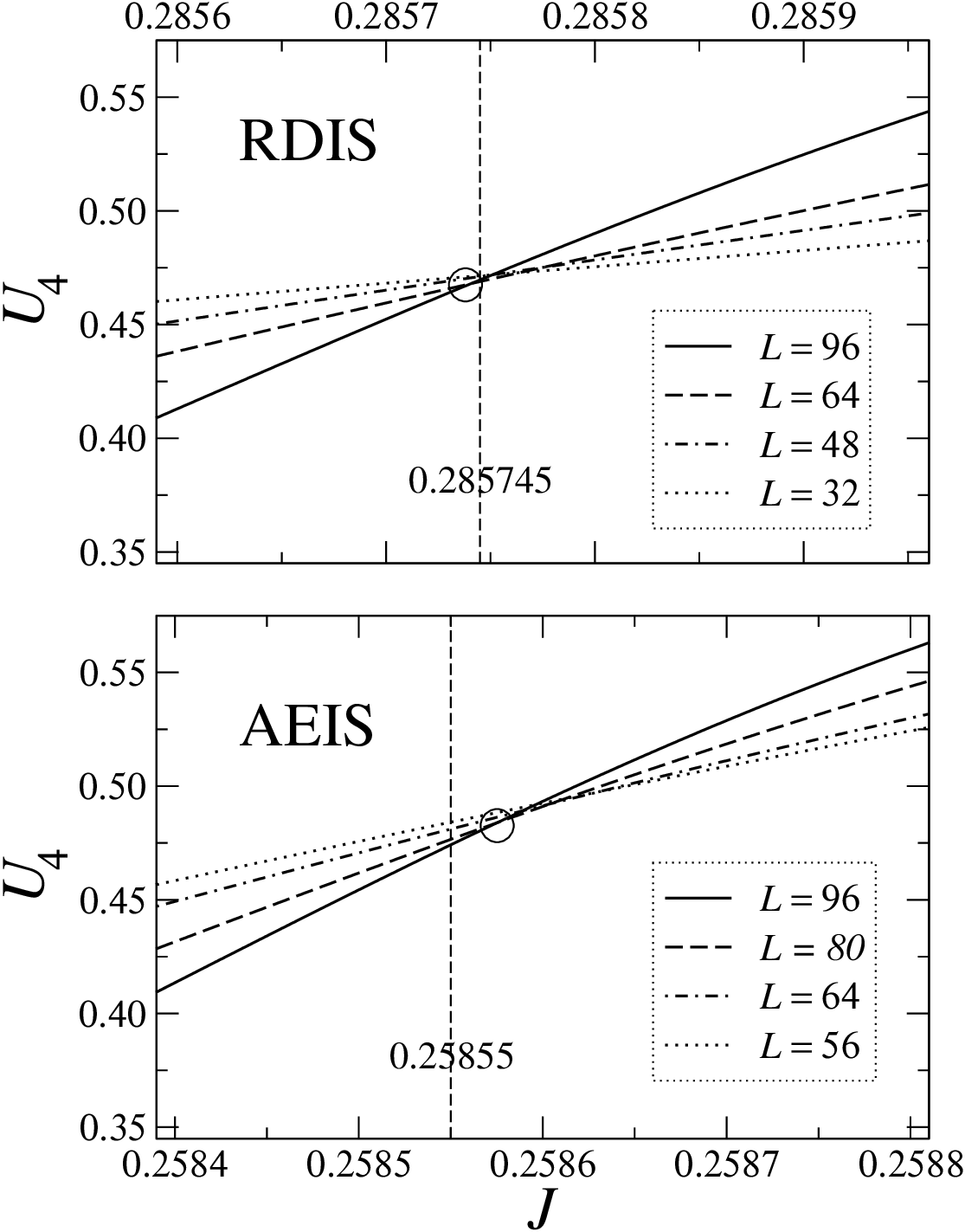}
\caption{Binder magnetization fourth cumulant as a function of $J$ for the RDIS 
({\em top}) and the AEIS ({\em bottom}), using the largest $L$. Dashed vertical lines 
indicate couplings $\widetilde{J_c}$ used in simulations, empty circles are 
intersections between curves at the two largest sizes. In the AEIS case, $J_c$ from 
this intersection and the previous estimate $\widetilde{J_c}$ differ by 
$\Delta J_c/J_c\approx10^{-4}$.\label{VMvsbeta}} 
\end{figure}

We use two methods to estimate the critical coupling out of our present simulations. 
Binder magnetization fourth cumulant  
\be
U_4 = 1 - \frac{\langle M\rangle^4}{3\langle M^2\rangle^2}\label{cumuM}
\ee
is universal, {\em i.~e.}, independent on the system size \cite{binder1} at the critical 
point. Thus, the critical coupling $J_c$ can be obtained with high accuracy at the 
point where $U_4$--$J$ plots coincide for all system sizes $L$. Fig.~\ref{VMvsbeta} 
shows these plots for the largest system sizes, for the RDIS (top) and the AEIS 
(bottom). In both cases, the coupling used in simulations has been marked by a vertical 
dashed line. For the RDIS case, taking the curves for the two largest sizes 
($L=64,96$), we obtain an estimate (circle) $J_c^{\tiny RDIS}=0.2857471(11)$ for the 
critical coupling, which agrees well with that of Calabrese 
{\em et al.}~\cite{calabreseMC}, $J_c^{RDIS}=0.2857447(24)$. In the AEIS case, we depict 
the same plots for lattice sizes $L=56-96$. The intersection for $L=80,96$ (circle) 
gives the critical coupling $J_c^{AEIS}=0.258575(10)$, close to the value used in 
simulations, $J_c^{AEIS}=0.25855$ ($\Delta J_c/J_c\approx 10^{-4}$). 

\begin{figure}[ht!]
\includegraphics[width=8cm]{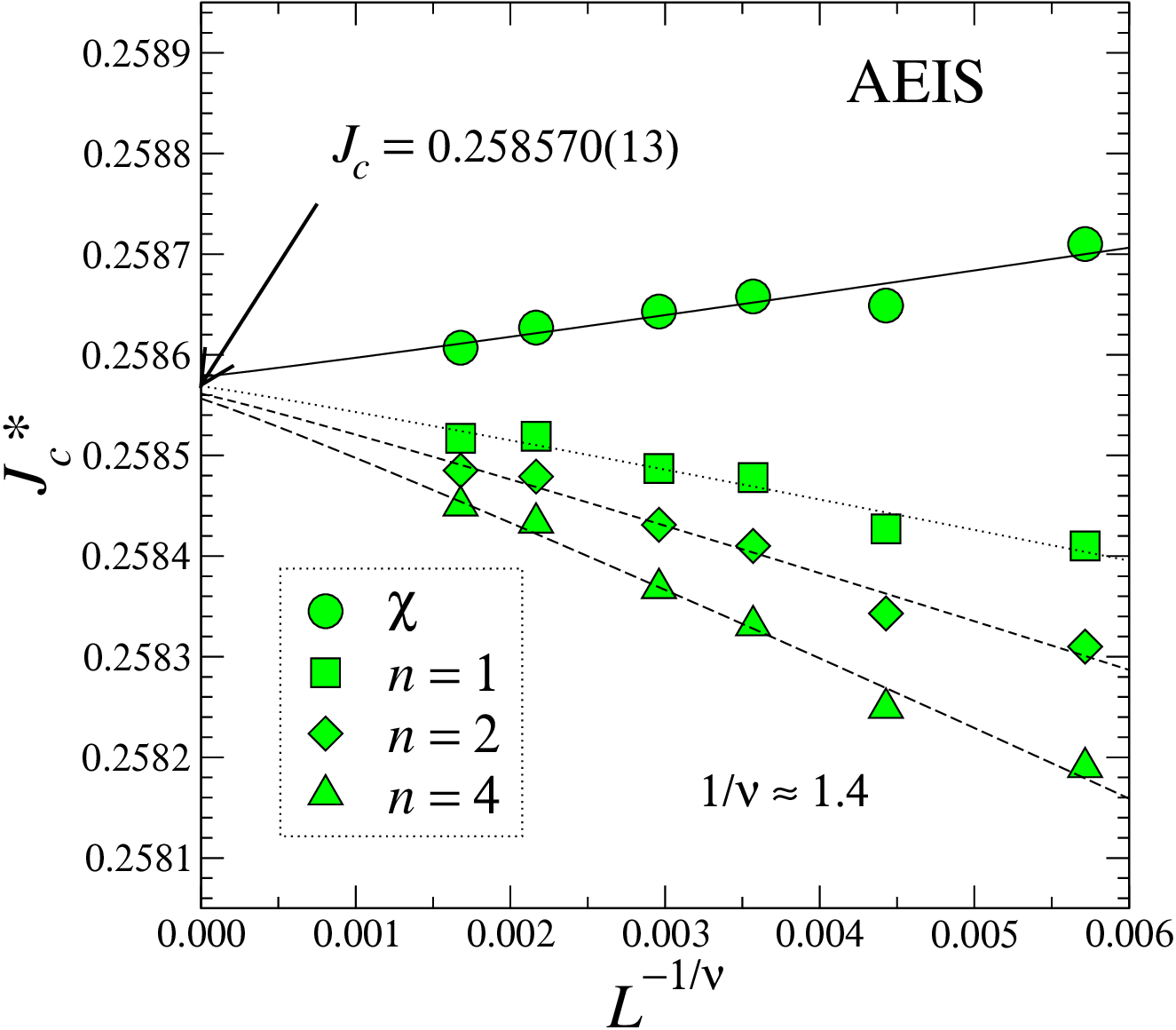}
\caption{(Color online) Position of the maxima $J^*_c$ for the
susceptibility and logarithmic derivatives of $M^n$ ($n=1,2,4$), plotted versus 
$L^{-1/\nu}$ using $1/\nu=1.4$. Linear fits average at $J=0.258570(13)$ 
in the thermodynamic limit ($L\to\infty$). \label{betamaxvsL}}
\end{figure}

On the other hand, following the FSS theory\cite{binder1}, deviations of pseudocritical 
couplings $J^*_c(L)$ from the critical coupling $J_c$ scale as: 
\be\label{Jc}
J^*_c(L) - J_c \sim L^{-1/\nu}\,,
\ee
where $J^*_c(L)$ is defined as the positions of maxima for a given critical quantity, 
being $J_c\equiv J^*_c(L\to \infty)$. 
For instance, values $J^*_c(L)$ for the susceptibility (\ref{suscep}) and logarithmic 
derivatives (\ref{logderiv}), obtained from reweighted curves, have been depicted in 
Fig.~\ref{betamaxvsL} as functions of $L^{-1/\nu}$, where the rough value 
$1/\nu \approx 1.4$ has been estimated through non-linear fits of points corresponding 
to $L=56-96$. As expected, the linear behaviour (\ref{Jc}) is observed, and the lines 
cross the $L\to\infty$ axis at an average point $J_c= 0.258570(13)$, quite close to the 
value estimated above, using the intersection of Binder fourth cumulants. 

\section{Effective exponents}\label{effective}

\subsection{Correlation length exponent}
Finite size scaling (FSS) \cite{fisher} has been used to estimate effective critical 
exponents for the RDIS and the AEIS. This method allows us estimate critical exponents 
$\beta/\nu$, $\gamma/\nu$, $\alpha/\nu$, and the correlation length inverse exponent 
$1/\nu$. The latter has already been roughly estimated above, using the scaling law for 
the position of the maxima of logarithmic derivatives of the magnetization moments 
\cite{ferrenberg} $\langle\mathcal{M}^n\rangle$ ($n=1,2,4$), and the susceptibility. 
More accurate estimations are made directly taking averages over disorder on quantities 
obtained at $\widetilde{J_c}=0.25855$, used in our extensive simulations, which is 
close to the previously estimated $J_c$: 
\be 
\left[ 
\frac{\partial\ln \langle\mathcal{M}^n\rangle}{\partial J} 
\right]_{J=\widetilde{J_c}} \sim \,L^{1/\nu}\,.\label{edlnMkvsL} 
\ee 
We look first to this exponent in order to determine the effect of disorder on the 
critical behaviour of the 3DIS. Previous works \cite{peberche} report that exponents 
$\beta/\nu$ and $\gamma/\nu$ for the RDIS are almost the same as those for the pure 
3DIS, and as shown later, this is the case for the AEIS. 

\begin{figure}[ht!]
\includegraphics[width=7cm]{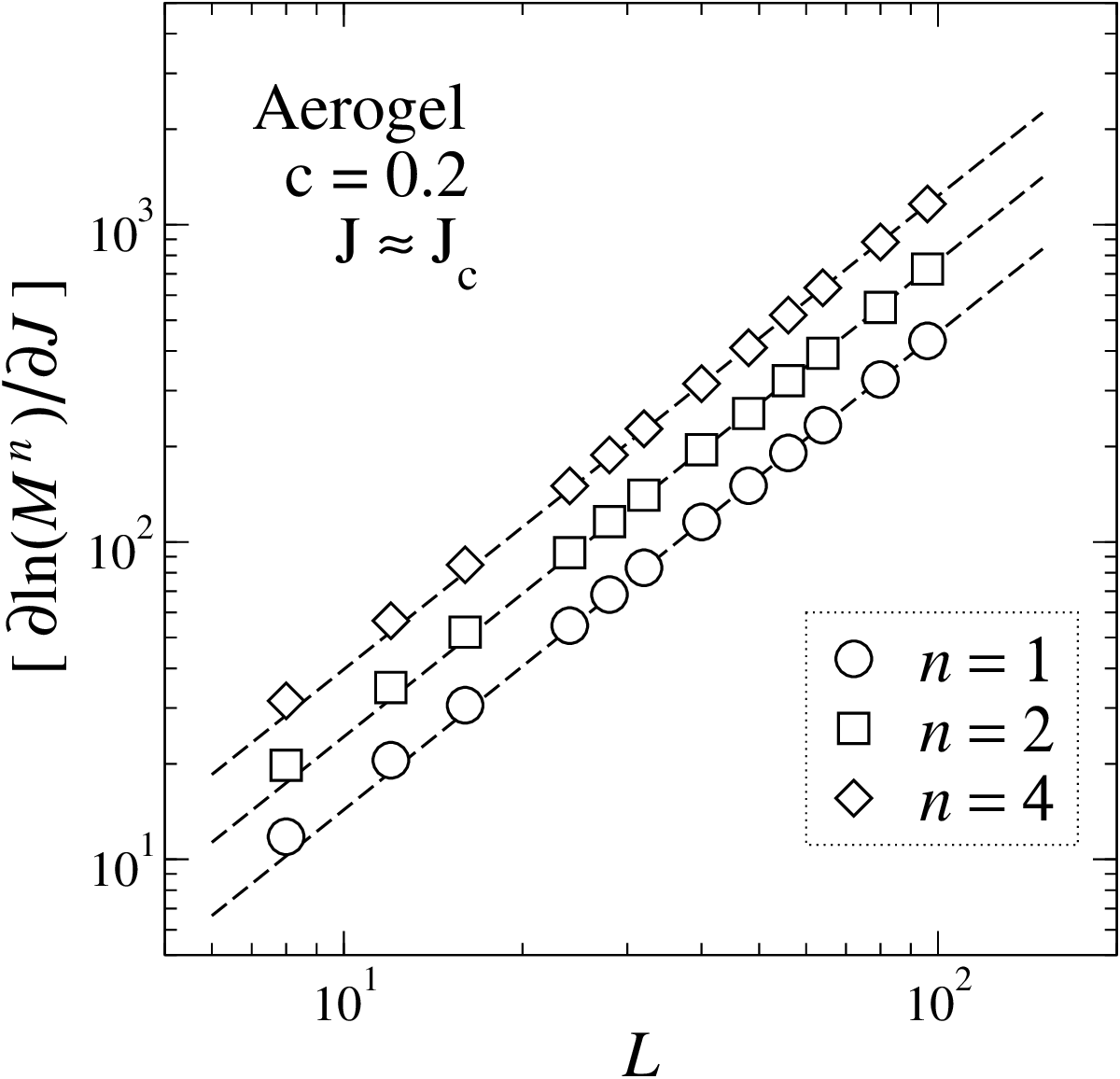} 
\caption{Logarithmic derivatives 
$\left[\partial\ln \langle\mathcal{M}^n\rangle/\partial J\right]$ ($n = 1, 2, 4$) 
versus $L$. Points are averages over disorder taken from our simulations at the 
estimate $\widetilde{J_c}=0.25855$ for the AEIS. Power-law fits (dashed lines) out of 
the four largest lattice sizes $L=56-96$ give a FSS exponent 
$1/\nu = 1.501\pm 0.007$. \label{dlnMkvsL}} 
\end{figure}

Logarithmic derivatives of moments $n=1,2,4$ of the magnetization are plotted, versus 
system size $L$, in Fig.~\ref{dlnMkvsL} for the AEIS case. Points were obtained from 
averages over disoder at the simulation coupling $\widetilde{J_c}=0.25855$ which is 
quite close to $J_c$, as estimated above. Dashed lines are power-law fits to the 
equation (\ref{edlnMkvsL}) using the four largest system sizes ($L=56-96$), and give a 
FSS exponent $1/\nu = 1.501\pm0.007$ for the AEIS. 
We recall that exponents determined by this method are effective exponents, and only 
their asymptotic behaviour would give a hint to what the universal critical exponent 
tend to. This study is addressed in next paragraphs. 

\begin{figure}[th!]
\includegraphics[width=7cm]{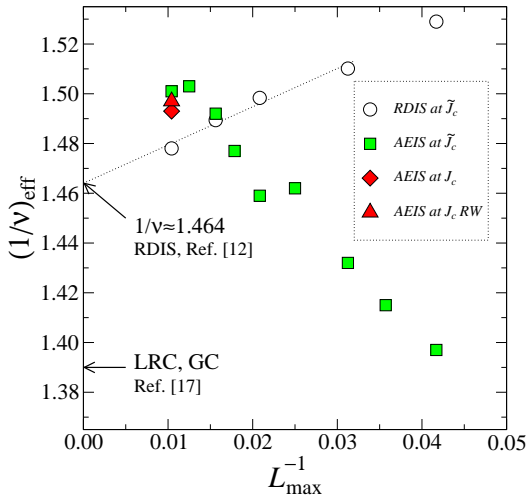}
\caption{(Color online) 
Effective correlation lenght inverse exponent $(1/\nu)_{\mbox{\small eff}}$ versus 
$1/L_{max}$, obtained by the FSS power-law fit (\ref{edlnMkvsL}) of 
four consecutive points ending at $L_{max}$. 
Values for the RDIS (empty circles), extrapolated to $L_{max}\to \infty$ 
(dotted line) approach the result by Calabrese {\em et al.}~\cite{calabreseMC}. 
In the AEIS case (filled squares), points clearly depart from the LRC fixed point 
through a region close but above the SRC fixed point. Averages over disorder on less 
extensive simulations of the AEIS at the critical coupling $J_c=0.258570$, give 
$(1/\nu)_{\mbox{\small eff}}\simeq 1.493$ (diamond). The value 
$(1/\nu)_{\mbox{\small eff}}\simeq 1.497$ (triangle) is obtained from averaged 
reweighted curves at $J_c=0.258570$.\label{nuvsLmax}}
\end{figure}

Effective exponents $(1/\nu)_{\mbox{\small eff}}$ have been depicted in 
Fig.~\ref{nuvsLmax} as calculated from FSS of logarithmic derivatives in both cases, 
RDIS and AEIS. As in Fig.~\ref{dlnMkvsL}, values were obtained from averages over 
disorder at the simulated couplings $\widetilde{J_c}\approx J_c$, being 
$\widetilde{J_c}=0.285745$ for the former, and $\widetilde{J_c}=0.25855$ for the 
latter, as stated above. Each value $(1/\nu)_{\mbox{\small eff}}$ is then obtained from 
power-law fits to the FSS expression (\ref{edlnMkvsL}), taking four consecutive points 
whose maximum size is $L=L_{max}$. 
Results for the RDIS (empty circles) yield $(1/\nu)_{\mbox{\small eff}}=1.478(5)$ at 
$L_{max} = 96$. A rough estimate of the asymptote $1/\nu$, is obtained 
by extrapolating these points to the $L_{max}^{-1}\to 0$ axis, as seen 
in Fig.~\ref{nuvsLmax} (dotted line). The extrapolation yields $1/\nu\approx 1.464$, 
well in agreement with previously reported results for the RDIS\cite{calabreseMC}. 
Results for the AEIS (filled squares) give $(1/\nu)_{\mbox{\small eff}}= 1.501(7)$ at 
$L_{max}=96$. Effective exponents in this case clearly depart from 
values corresponding to the LRC fixed point\cite{prudnikov}, through a region close 
but above the SRC fixed point at $L_{max}\approx 48$. However, at larger 
lattice sizes, greater values suggest that a another fixed point may rule the critical 
behaviour at the thermodynamic limit. 

For the stable uncorrelated (SRC) disorder fixed point, the theory \cite{Harris} 
predicts that the exponent $1/\nu$ should be smaller than $3/2$. Additionally, the WH 
condition \cite{wh83} (\ref{whlrc}) is well satisfied for this AEIS, where $\nu$ is 
the pure 3DIS exponent, and $a=2(d-d_f)$ comes from LRC of gelling clusters (GC) within 
DLCA aerogels\cite{vasquez}. In effect, the fractal dimension for the GC within 
aerogels at $c=0.2$ is $d_f\approx 2.2$, as reported elsewhere\cite{vasquez,vasquez3}. 
This condition, together with theoretical predictions reported by Prudnikov 
{\em et al.}~\cite{prudnikov}, would give $1/\nu\approx 1.4$ for the 3DIS with LRC 
defects, at the corresponding $a\approx 1.6$. From Fig.~\ref{nuvsLmax} it is clear that 
$(1/\nu)_{\mbox{\small eff}}$ is far above this value. Thus, it is not the LRC subset 
of disorder (the GC) which rules the critical behaviour of the AEIS, in the way it 
certainly does for 3DXY universality class in aerogels\cite{vasquez}. 

The exponent $(1/\nu)_{\mbox{\small eff}}$ at $L_{max}=96$ was also obtained taking 
averages of logarithmic derivatives from average reweighted curves, at the critical 
coupling estimated above (Section \ref{secJc}), $J_c=0.258570$. For $L=56-96$ we obtain 
$(1/\nu)_{\mbox{\small eff}}\approx 1.497$ (triangle, Fig.~\ref{nuvsLmax}). 
In addition, less extensive additional realizations of the AEIS also for $L=56-96$ 
($600$ for each size), were made at this more accurate value $J_c=0.258570$, and 
averages over disorder were taken directly from simulations. The power-law fit for 
these points gives an estimate $(1/\nu)_{\mbox{\small eff}}\approx 1.493$ (diamond). 
Although these effective exponents are lower than $3/2$, there exists yet not enough 
evidence in this work that the RDIS fixed point would be reached at 
$L_{max}\to \infty$. 

\subsection{Specific heat and energy exponents}
To check our results about the correlation length exponent for the AEIS, we study the 
FSS of the specific heat and the energy, at the simulation coupling $\widetilde{J_c}$.
\begin{figure}[ht!]
\includegraphics[width=8cm]{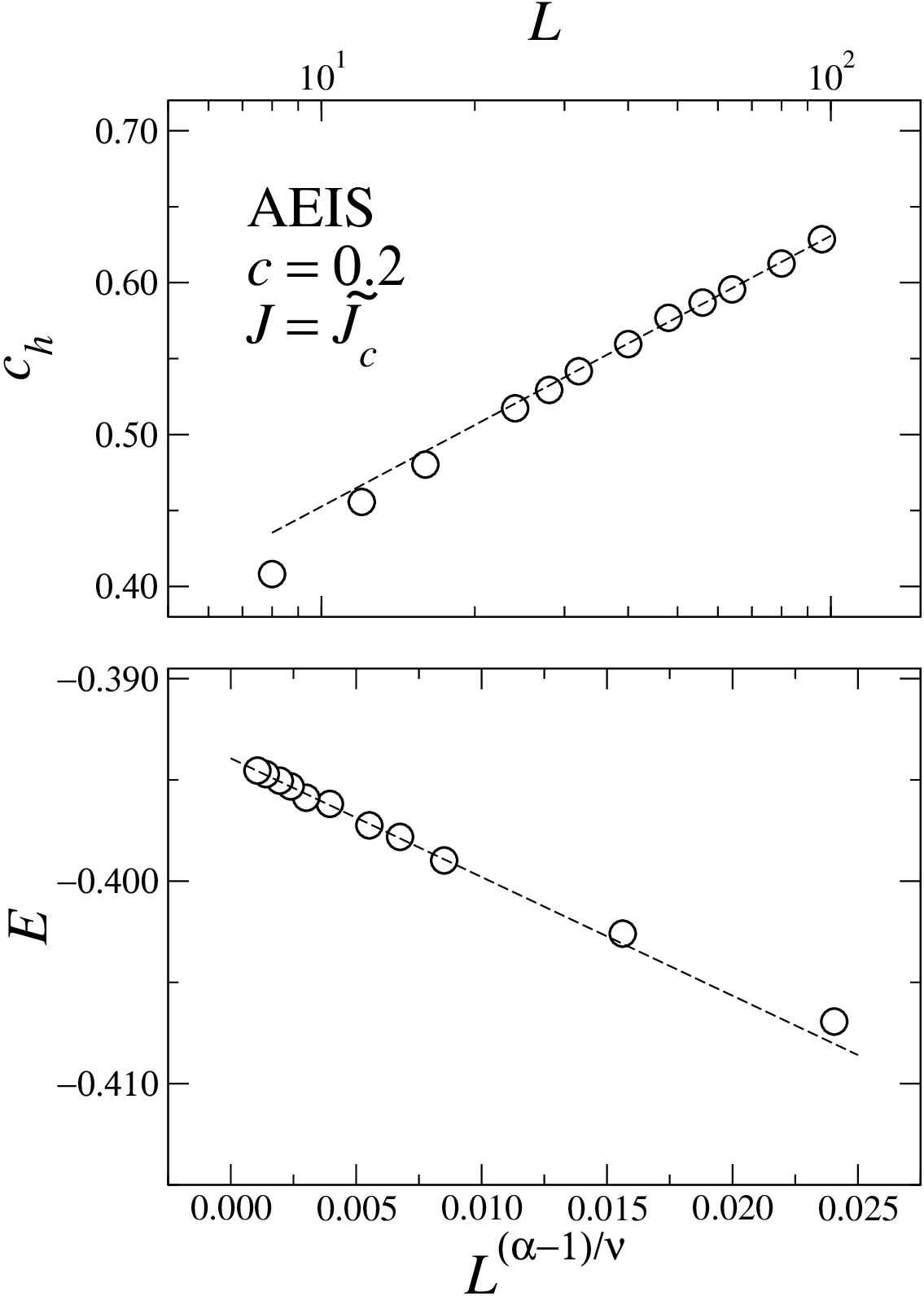}
\caption{
({\em Top}) Specific heat $c_h$  versus $L$ for the AEIS in a linear-log scale. A 
logarithmic singularity ($\alpha=0$) for the specific heat at $\widetilde{J_c}$ follows 
from the fit (dashed line). This result reinforces the estimated $1/\nu\approx 1.5$ 
(Fig.\ref{dlnMkvsL}). ({\em Bottom}) Energy plotted against  $L^{(\alpha-1)/\nu}$, taking 
$\alpha=0$ and $1/\nu=3/2$, and the corresponding linear fit is shown (dashed line).
\label{cv_EvsL}
}
\end{figure}
 
Top of Fig.\ref{cv_EvsL} shows the specific heat $c_h$ plotted versus $L$ using a 
linear-log scale. The dashed line is a logarithmic fit using the four largest lattice 
sizes, $56-96$. This results suggests that the singularity of the specific heat could 
be logarithmic, consistent with our result $1/\nu \approx 1.5$ for $L_{max}=96$. 
After Josephson hyperscaling relation ($\alpha= 2 - d\nu$) a specific heat exponent 
$\alpha \approx 0$ would be expected. Bottom of Fig.\ref{cv_EvsL} shows the linear 
dependence of the energy $E$ on $L^{(\alpha-1)/\nu}$, taking $\nu=2/3$ and $\alpha=0$, 
which confirms the results stated above. 
We made additional analisys to specific heat data, and the energy as well, using 
the scaling of both quantities in the case $\alpha/\nu<0$. This method was performed 
by Schultka and Manousakis in determining the (negative, very small) exponent 
$\alpha/\nu$ for the pure 3DXY model \cite{manousakis}. In the case $\alpha/\nu<0$, 
the specific heat scales as $c_h = c_{\infty} + c_1L^{\alpha/\nu}$, while the energy 
scales as $E=E_{\infty} + E_1L^{(\alpha-1)/\nu}$. Non-linear fits to these expressions, 
using the six largest lattice sizes, $40-96$, give self-consistent results 
$\alpha/\nu \approx -0.022$ and $(\alpha-1)/\nu\approx-1.512$, in agreement with a 
correlation lenght exponent $1/\nu\approx1.49$. This result agrees with the tendency 
shown by the effective values in Fig.\ref{nuvsLmax}. Experiments on the critical point 
of the LV transition of $^4$He in aerogels \cite{wong4He} report a cusplike peak in the 
specific heat, but authors do not report an esimate for the exponent $\alpha$. Our 
results point for the largest lattice sizes to a logarithmic singularity, which may 
also be consistent with these experimental results. 

\begin{figure}[ht!]
\includegraphics[width=8cm]{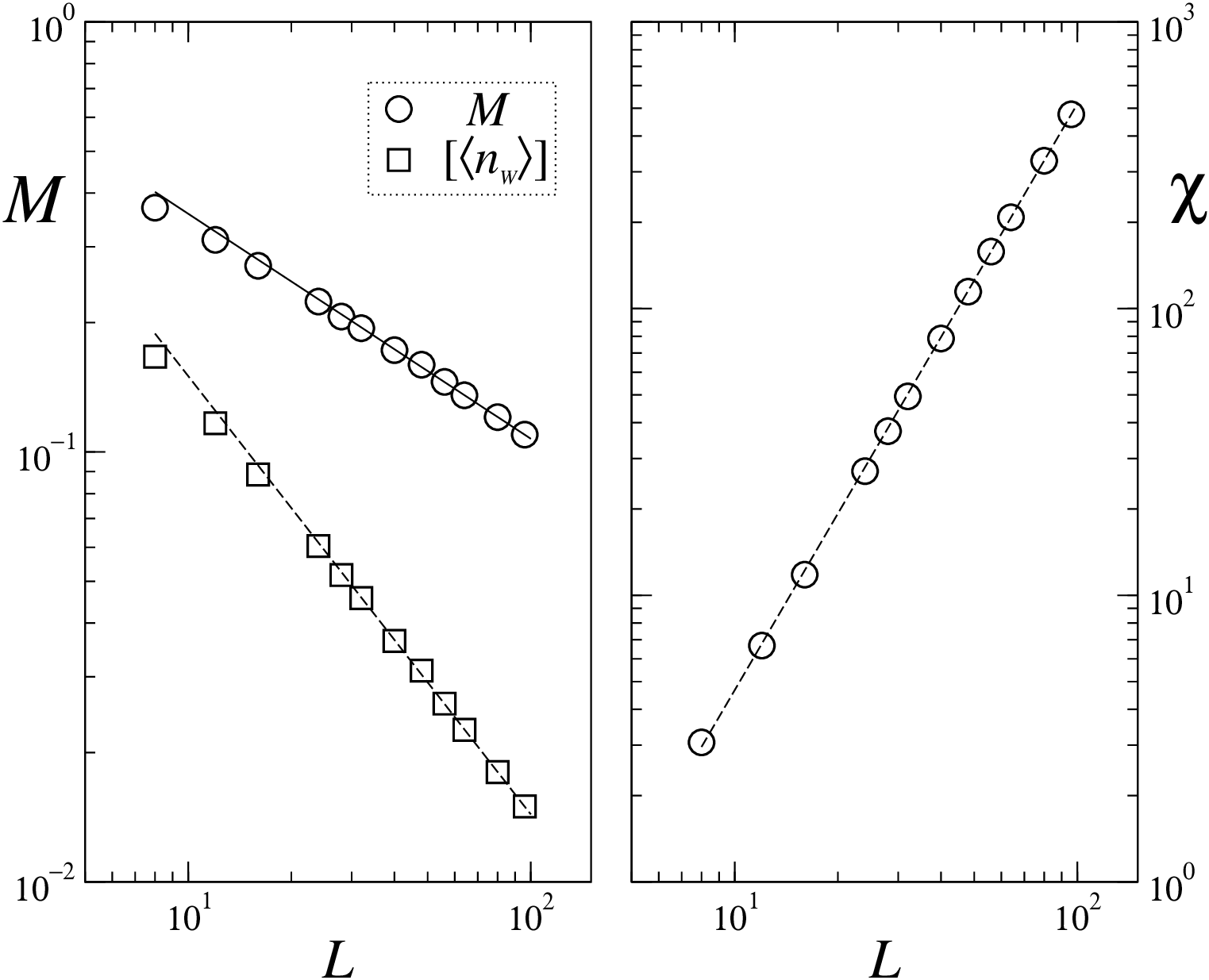}
\caption{Order parameter ({\em left}) and susceptibility ({\em right}), at the simulated 
critical point $\widetilde{J_c}$, versus $L$ for the AEIS. The magnetization $M$ 
(circles) scales as $L^{-\beta/\nu}$, and the averaged Wolff cluster size 
$\left[{\langle n_{\mbox{\tiny W}}\rangle}\right]$ scales as \cite{Wolff2} 
$L^{-2\beta/\nu}$ (squares), giving an average estimate of 
$\overline{2\beta/\nu}= 1.032(1)$. The FSS power-law fit for the susceptibility gives 
$\gamma/\nu= 2.044(4)$. All power-law fits have been made for $L=56-96$.\label{X,MvsL}} 
\end{figure}
 
\subsection{Magnetic exponents} 
According to the FSS theory, the magnetization and the susceptibility scale as 
$M \sim L^{-\beta/\nu}$ and $\chi\sim L^{\gamma/\nu}$, respectively.
In Fig.~\ref{X,MvsL}, we plot the order parameter $M$ ({\em left}) and the 
susceptibility ({\em right}) as a function of $L$ for the AEIS. Magnetization data 
(circles) have been fit to the preceding power-law FSS expression (continuous line), 
givieng $\beta/\nu= 0.523(3)$. Average sizes of Wolff clusters divided by $L^3$, 
$\left[{\langle n_{\mbox{\tiny W}}\rangle}\right]$ (squares), scale with the same 
exponent as the squared magnetization\cite{Wolff2}. This is confirmed by the power-law 
fit (dashed line) which yields $2\beta/\nu=1.019(6)$. These results give an average 
estimate $\overline{2\beta/\nu}= 1.032(6)$. Together with $1/\nu\approx 1.5$ this gives 
$\beta= 0.34(4)$, close to the the pure 3DIS exponent and to the RDIS exponent. On the 
right side of Fig.~\ref{X,MvsL}, points for the susceptibility obtained from 
simulations near the critical point fit to the FSS power-law expression with the 
exponent $\gamma/\nu= 2.044(4)$. All fits have been made for $L=56-96$. 

\begin{figure}[ht!]
\hspace{-0.5cm}
\includegraphics[width=6.5cm]{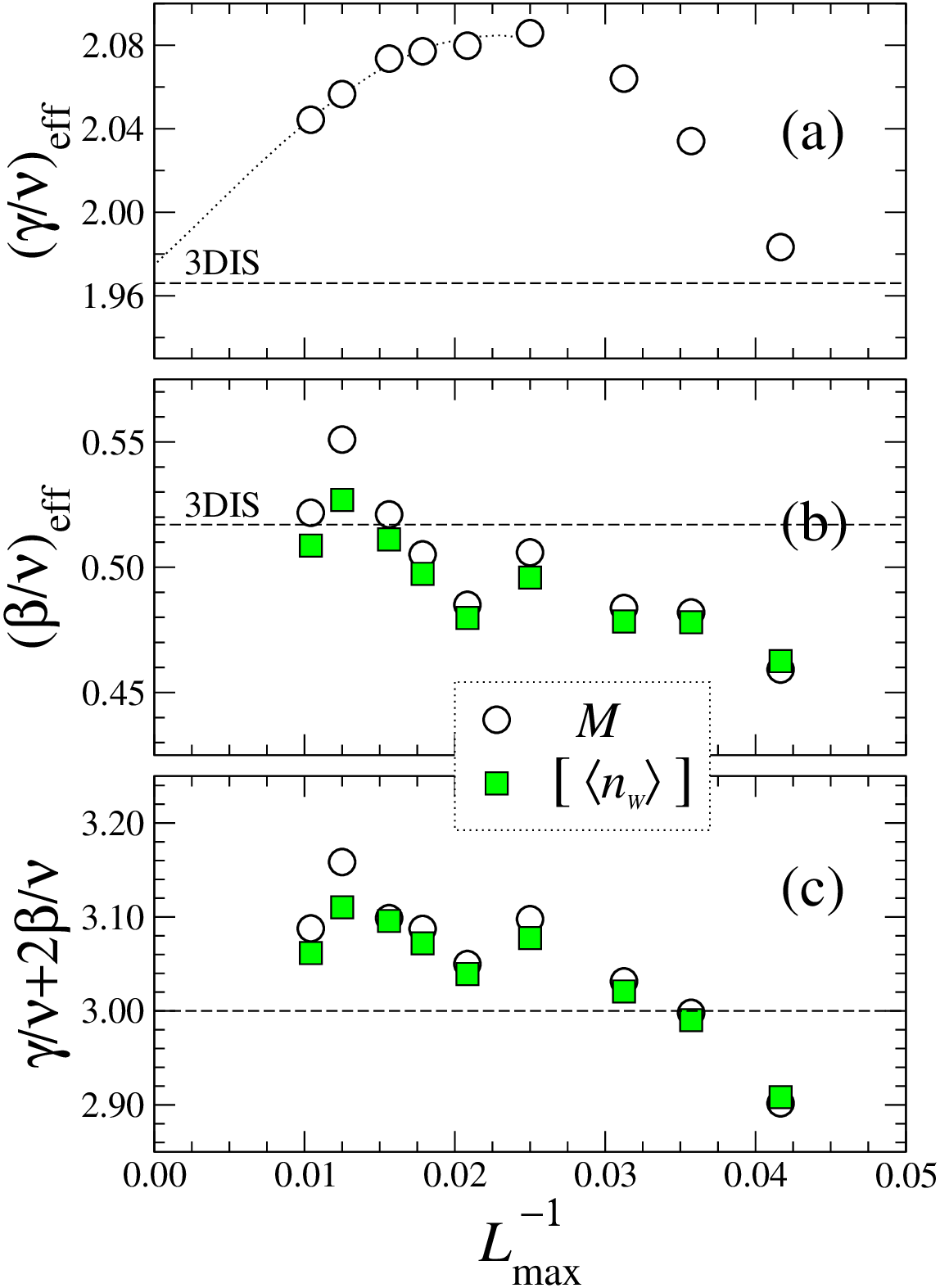}
\caption{(Color online) (a) Effective exponents for the susceptibility give 
$(\gamma/\nu)_{\mbox{\small eff}}=2-\eta_{\mbox{\small eff}} > 2$. 
Extrapolation using $L_{max}\geq 48$ suggest that $\gamma/\nu<2$ 
in the thermodynamic limit. (b) Effective exponents $(\beta/\nu)_{\mbox{\small eff}}$ 
for the the magnetization (circles) and for the average mass of Wolff clusters 
(squares). At $L_{max}=96$ both approach the pure 3DIS theoretical estimate\cite{Guida} 
$(\beta/\nu)_{\mbox{\small eff}}\approx 0.517$. (c) The hyperscaling relation 
$\gamma/\nu+2\beta/\nu=d=3$, not satisfied for effective exponents, tends to hold for 
larger lattice sizes.\label{expovscaja}} 
\end{figure} 
Results for magnetic effective FSS exponents, obtained by fitting four consecutive 
points from Fig.~\ref{X,MvsL}, ending at $L_{max}$, have been depicted in 
Fig.~\ref{expovscaja}. Fig.~\ref{expovscaja}(a), shows the effective exponent for the 
susceptibility, $(\gamma/\nu)_{\mbox{\small eff}}$, plotted versus $L_{max}^{-1}$. This 
exponent tends to increase for $L_{max}<48$, but beyond this size the tendency is to 
stabilize at a value close to that of percolation, $\gamma/\nu=2-\eta\approx 2.05$. For 
larger $L_{max}$, it turns to approach a value lower than $2.00$ (positive $\eta$). The 
asymptotic extrapolated value seems to be the pure 3DIS exponent $\gamma/\nu=1.966(3)$ 
\cite{Guida} or the RDIS \cite{ballesteros} $\gamma/\nu=1.963(5)$. Berche and 
collaborators \cite{peberche} estimated effective values $\gamma/\nu > 2$ for smaller 
concentrations in the RDIS case. Using the result for $L_{max}=96$, 
$\gamma/\nu=2.044(4)$, and our estimate $\nu=2/3$, the exponent $\gamma$ found in the 
present work is $1.363(9)$, slightly above the value $\gamma=1.344(9)$ found in our MC 
simulations for the RDIS, and $\gamma=1.342$ obtained by Calabrese 
{\em et al.}~\cite{calabreseMC} for the same system. The tendency for larger $L$ is to be 
closer to RDIS results. 

We must remind at this point, that theoretical most accurate results, by Prudnikov 
{\em et al.}~\cite{prudnikov}, predict a magnetic exponent $\eta<0$ for the 3DIS in LRC 
disordered structures with an algebraic decay similar to that of the gelling clusters 
within DLCA aerogels at $c=0.2$ \cite{vasquez3}. 

On Fig.~\ref{expovscaja} (b) and (c), squares represent effective exponents 
$(\beta/\nu)_{\mbox{\small{eff}}}$ obtained from 
$\left[{\langle n_{\mbox{\tiny W}}\rangle}\right]$, and circles, those obtained 
directly from $M$. There is a strong variation of these results with $L_{max}$. As 
stated above, the exponent obtained averaging both results using $L_{max}=96$ is 
$\overline{\beta/\nu}= 0.516(6)$, close to that of the pure 3DIS $\beta/\nu=0.517(3)$ 
\cite{Guida}. In addition, our results agree well with those reported for the RDIS case 
by Ballesteros {\em et al.}~\cite{ballesteros}, $\beta/\nu=0.519(3)$. In this work, we 
obtain $\beta/\nu=0.516(5)$ for the RDIS. The last effective value 
$(\beta/\nu)_{\mbox{\small{eff}}}$ ($L_{max}=96$), together with $1/\nu\approx 1.5$ 
gives an exponent $\beta=0.343(9)$ for the order parameter. This result agrees well 
with experiments about the critical point of the LV transition of $N_2$ in $95\%$ 
porous aerogels ($c=0.05$), reported by Wong {\em et al.}~\cite{wongN2}, which yield 
$\beta = 0.35(5)$. 

Care must be taken with this agreement because, our results were obtained using DLCA at 
concentrations $c=0.2$, and these structures are quite different from those at 
$c=0.05$. For the latter, most of impurities belong to the LRC gelling clusters, giving 
the DLCA aerogels a less random overall structure. 

As stated above, a possible explanation for these magnetic exponents is the influence 
of the LRC disorder fixed point. The fractal dimension of the aerogel gelling cluster 
is $d_f\approx 2.2$, giving an exponent $a\approx 1.6$ associated to this structure 
\cite{vasquez}. Following Table IV from Prudnikov {\em et al.}~\cite{prudnikov}, a value 
$\gamma/\nu > 2.0205$ ($\eta<-0.0205$) is expected. The effective value found in this 
work is close to this prediction, but it follows from Fig.~\ref{expovscaja}(b) that a 
tendency exists to approach a value closer to the corresponding RDIS fixed point. 
Finally, using effective values the hyperscaling relation $\gamma/\nu+2\beta/\nu=3$ 
seems not to hold, as seen in Fig.~\ref{expovscaja}(c). 
The violation of this hyperscaling relation suggest that our results do not yet 
reach asymptotic values. Extensive simulations rest still to be performed at the more accurate value 
$J_c=0.258570$. 

\section{Concluding remarks}\label{conclu}
Extensive Monte Carlo simulations of the 3D Ising model (3DIS) with impurities have 
been reported in this paper. Using finite size scaling, critical couplings and 
exponents have been estimated for the 3DIS, in presence of randomly distributed 
impurities (RDIS), and confined in aerogel-like structures (AEIS). For the latter we 
have collocated Ising spins in the pores of simulated DLCA aerogels at $c=0.2$. At this 
concentration, this objects are known to be non-fractal. However, the presence of 
{\em hidden} LRC could affect criticality, as predicted by the theory\cite{prudnikov}. 
It has been concluded elsewhere\cite{vasquez} that these LRC structures, the gellling 
clusters, modify the critical behaviour of the 3DXY model, when confined in the same 
kind of aerogel-like structures. In the 3DIS case, however, our results for thermal 
exponents $1/\nu\lesssim 1.5$ and $\alpha/\nu\lesssim 0$ rest far above those 
for the LRC fixed point predicted by the theory\cite{prudnikov}. Complementary 
simulations at a more accurate value of the critical coupling, $J_c=0.258570(13)$, 
give an exponent $1/\nu\approx 1.49$. Although similar thermal exponents have been 
reported by Pakhnin and Sokolov\cite{pakhnin} for the RDIS universality class, the 
asymptotic critical regime could have not been reached in our simulations, and more 
extensive simulations are yet to be performed at this more accurate $J_c$ value. 

Effective critical exponents observed here for the AEIS change from a fixed point (LRC) 
at box sizes $L\leq 48$, to another (SRC) at box sizes $L>48$ (Figs. \ref{nuvsLmax} and 
\ref{expovscaja}), probably indicating an oscillating approach to the stable fixed 
point. Theoretical predictions based on the Weinrib and Halperin model 
\cite{wh83,prudnikov}, able to explain changes on the critical behaviour of the 3DXY 
model in the pores of DLCA aerogels\cite{vasquez}, may also explain the influence of 
this type of disorder on the 3DIS. In this case, two competing effects are present: the 
random SRC subset of the disorder (defined in Section \ref{dlcaproc}), which already 
affects the critical behaviour of the 3DIS, and the LRC subset which, after the 
extended criterion (\ref{whlrc}), may be relevant as well. This is certainly not the 
case for the 3DXY model, where Harris criterion prevents the SRC subset of impurities 
(islands) to be relevant: only the weak LRC distribution of impurities (GC) is 
relevant\cite{vasquez} for the 3DXY. For the 3DIS, theory predicts that both, LRC and 
SRC subsets may be relevant. Which one finally dominates the critical behaviour? 

Results presented in this paper suggest that, in the AEIS case, the critical behaviour 
is ruled by the SRC fixed point. A plausible explanation to these dominating SRC 
effect is provided by theoretical works\cite{wh83,prudnikov}: for the 3DIS, RG flows 
converge to a more stable SRC fixed point, because at $m=1$ the LRC fixed point is less 
stable (marginal). However, it has been mentioned before, without a proof\cite{wh83}, 
that amplitudes of disorder may in some cases affect criticality. Added in proof, we 
have to mention that in preliminary simulations of the 3DIS in presence of mixed kinds 
(LRC and SRC) of disorder\cite{manuel}, evidence of a continuous flow from the LRC 
fixed point to the SRC one has been observed, when relative strengths are tunned from a 
pure LRC distribution of defects to a $1\!:\!1$ proportion. In the AEIS case, we have 
analized relative amplitudes (strengths) of the LRC and the SRC subsets of disorder, for $L=128$,to 
determine that up to $97\%$ of defects are due to islands (SRC), while only $3\%$ are 
due to the GC (LRC). 

To conclude, it has to be stated that the influence of aerogel-like distributions of 
impurities on the critical behaviour of the 3DIS is yet far from being completely 
understood. The problem is similar to that of phase transitions in Ising systems with 
non-integer dimension or in fractal structures. It becomes clear that the fractal 
dimension, related to the exponent $a$ of subjacent long-range correlations, may not be 
the only parameter to determine the universality class of the impure system. 
\cite{noninteger}. 

\section*{ACKNOWLEDGEMENT}\label{aknow}
Authors thank CNRS and FONACIT (PI2004000007) for their support. Discussions with 
N.~Olivi-Tran, B.~Berche, Yu.~Holovatch, and M.~Marqu\'es are kindly acknowledged. 
Invaluable discussions with A.~Hasmy and R.~Jullien have improved our understanding on 
aerogel structure. C.V. kindly expresses gratitude to all the personnel of the LCVN at 
Montpellier, France. 
\newline

\end{document}